# Forecasting election results by studying brand importance in online news

Fronzetti Colladon, A.

**Forecasting Election Results by Studying Brand Importance in Online News**

Fronzetti Colladon, A

**Abstract**

This study uses the Semantic Brand Score, a novel measure of brand importance on big textual data, to forecast elections based on online news. About 35,000 online news articles were transformed into networks of co-occurring words and analyzed combining methods and tools of Social Network Analysis and Text Mining. Forecasts, made for four voting events in Italy, provided consistent results across different voting systems: a general election, a referendum and a municipal election in two rounds. This work contributes to the research on electoral forecasting by focusing on predictions based on online big data; it offers new perspectives about the textual analysis of online news, through a methodology which is relatively fast and easy to apply. This study also suggests the existence of a link between brand importance of political candidates and parties and electoral results.

**Keywords:** Election forecasting; semantic brand score; social network analysis; text mining; online news; big data.

## 1. Introduction

There was a long tradition of research on electoral forecasting (Campbell, 1992; Lewis-Beck & Rice, 1984), with a large interest for this topic which persists nowadays (Lewis-Beck, 2005; Rothschild, 2015; Vergeer, 2013; Williams & Reade, 2016). An accurate nowcasting, or long-term forecasting, of these results is in the desires of a large set of stakeholders, such as politicians, practitioners and policy makers. Such achievement would drive strategic choices and influence political strategies, as well as the financing of political parties. So far, many complex models were developed with the purpose of making more accurate predictions (Linzer & Lewis-Beck, 2015) – with some of them based on the analysis of polls, political markets or big data. In this scenario, the Internet and the availability of online data sources began to play a major role, leading to more accurate predictions than those solely based on opinion polls (MacDonald & Mao, 2016). The Internet enhances the information available to voters and stimulates increased participation (Tolbert & McNeal, 2003).

Despite the availability of large datasets which include information about online news (Leban, Fortuna, Brank, & Grobelnik, 2014; Leetaru & Schrodt, 2013), scholars have often based their forecasts on the analysis of tweets (Burnap, Gibson, Sloan, Southern, & Williams, 2016; Ibrahim, Abdillah, Wicaksono, & Adriani, 2016), or of search-volume indexes such as Google Trends (Mavragani & Tsagarakis, 2016). However, in many contexts, online news represent a pivotal information source, whose analysis can support economic, financial and political predictions (Elshendy & Fronzetti Colladon, 2017; Garcia, Silva, & Correia, 2018). Indeed, only few people have personal contact with politicians and most voters get information about politics and public affairs from TV and online. A recent research of the Reuters Institute (Newman, Fletcher, Kalogeropoulos, Levy, & Nielsen, 2017) shows a clear generational divide, with people in the age range 18-44 mostly reading news online and older groups (45+) mostly watching TV newscasts. TV and online news seem to count much more than social media, whose role is often overplayed. These are additional findings which support the need for a deeper examination of online news and their connection with electoral outcomes. Accordingly, this paper presents an analysis of about 35,000 online news articles, which commented on four voting events in Italy: the constitutional referendum of 2016, the Rome Municipal Election of 2016 (first and second round) and the Italian General Election of 2018.

The Semantic Brand Score (SBS) – a novel measure of brand importance for big textual data (Fronzetti Colladon, 2018) – has been used to assess the relative strength of the names of political candidates, or of the terms Yes and No in the case of the referendum. The results, presented in Section 5, suggest the existence of an association between brand importance and election outcomes. Even if each event had a different voting system, the SBS could predict quite accurately the real vote in all cases, with an optimal lead time of one week.

This work represents a step forward in the analysis of big data collected from online news and offers new perspectives for the forecasting of electoral results; it also contributes to the research on brand importance, presenting a new use of the SBS, a recent and promising indicator whose potentialities are largely unexplored. The forecasting approach presented in this paper is useful for extracting information from online news, with a novel tool which allows the quantification of brand importance. This approach can partially solve some of the challenges faced by existing forecasting methods, like polls, which have a larger cost of implementation (in terms of time and money) and might be affected in the case of declining response rates or, for example, when nonvoters are under-sampled (Mellon & Prosser, 2017). The SBS tool can be used repeatedly at practically no cost, with the possibility of obtaining almost real time forecasts; this timely results can also be used to integrate information available in the gap between subsequent polls – or in the timeframes when the publication of poll results is prohibited by the law.

The author's findings suggest the existence of a strong link between brand importance in online news and votes received by political forces. Online news are not only important for forecasting purposes, but should be carefully considered as a major communication channel to convey political messages and influence voting intentions (Garcia et al., 2018; Tolbert & McNeal, 2003).

## 2. Conceptualization and Measurement of Brand Importance

In this study, brand importance is evaluated through the Semantic Brand Score, i.e. following the conceptualization proposed by Fronzetti Colladon (2018), which is summarized in this section. The SBS comprises three dimensions – brand prevalence, diversity and connectivity – each contributing to the general construct. These dimensions are tightly related to well-known models of brand knowledge and equity (Keller, 1993; Wood, 2000). Prevalence captures the frequency of appearance of a brand name in a discourse, it represents its visibility and it is related with the dimension of brand awareness (Aaker, 1996; Keller, 1993). The more a brand name is mentioned in a text document the higher the visibility and the probability that it will be recalled and recognized by the reader. Higher prevalence also suggests that the brand name is familiar to the text author. Diversity is a measure of the heterogeneity of the words co-occurring with a brand, partially related to the constructs of brand image and brand associations (Keller, 1993; Wood, 2000). If a brand is frequently mentioned, it has a high prevalence; however, it might be the case that the brand is mentioned always together with the same set of words. In such a scenario, diversity would be low, as the discourse about the brand would be limited to a very specific context. Usually brands with more heterogeneous associations are preferred, as embedded in a richer discourse (Fronzetti Colladon, 2018). This is also consistent with past research which showed that a higher number of associations is beneficial to brand strength (Grohs, Raies, Koll, & Mühlbacher, 2016). The third dimension, connectivity, is a new construct which expresses the level of 'embeddedness' of a brand name in a discourse (or set of articles, or other text documents). "One could imagine connectivity as a measure of 'how much of a discourse passes through a brand', i.e. how much a brand can support a connection between words which are not directly co-occurring. […] connectivity could be considered as the 'brokerage' power of a brand" (Fronzetti Colladon, 2018, p. 152), i.e. its ability to link different discourse topics. If a brand is important only for a small part of a broad discourse, it could be that its prevalence and diversity are high, but its connectivity is low. Past research also suggested that connectivity of a brand name in online semantic networks is linked to brand popularity and could support the prediction of firm financial performance (Fronzetti Colladon & Scettri, 2019; Gloor, Krauss, Nann, Fischbach, & Schoder, 2009). In summary, the differential response to knowledge of a brand name theorized by Keller (1993), is at least partially captured by the SBS – by prevalence, diversity and connectivity, which measure the brand name frequency of use, the heterogeneity of its textual associations and its embeddedness at the core of a discourse.

All the above-mentioned metrics are derived from social network analysis and text mining and their calculation is better described in Section 4. In general, they should not be intended as a strict replacement of past brand equity models. This new conceptualization of brand importance adapted and extended widely accepted models (Keller, 1993; Wood, 2000) to provide an assessment tool suitable for big textual data.

The use of the SBS is not restricted to company or product names: this metric can be applied to any word or set of words which appear in a text corpora. In the case studies presented in this paper, the names of political candidates and party leaders are taken as brands to evaluate their importance in online news and see if it can be used to anticipate election outcomes.

## 3. Forecasting Elections

The forecasting of electoral results has attracted the interest of many scholars, who developed several models, based on polls, expert judgements, or the analysis of other data sources (Graefe, Armstrong, Jones, & Cuzán, 2014; Linzer & Lewis-Beck, 2015). Among traditional approaches one could list: political stock markets, where forecasts are provided by looking at the candidates' investment shares; vote expression surveys, where voters are asked who they will vote for; vote expectation surveys, where the interviewees share their prospect about the winner of an upcoming election, regardless of their preferences; and statistical models (Lewis-Beck, 2005). Different methodologies can lead to varying forecasting performances, also depending on the voting system and country analyzed. While some scholars were more confident about the potentialities of statistical or integrated models (Lewis-Beck, 2005; Lewis-Beck & Dassonneville, 2015; Magalhães, Aguiar-Conraria, & Lewis-Beck, 2012), others favored the outcomes produced by polls (Graefe, 2015), or the analysis of prediction markets (Williams & Reade, 2016). However, these models are not free from errors. Many existing approaches "overlook the uncertainty in coefficient estimates, decisions about model specifications, and the translation from popular vote shares to Electoral College outcomes" (Lauderdale & Linzer, 2014, p. 965).

In general, forecasts based on the analysis of single or aggregated opinion polls had an alternating success, with forecasting errors often varying in a larger range than those obtained by political science modelers (Tien & Lewis-Beck, 2016). Accuracy of vote intention polls, as the one of other models, is influenced by the characteristics of political systems, and errors are usually larger for greater parties (Jennings & Wlezien, 2018). Average poll errors usually decrease over time, with more informative results appearing in the last days – sometimes excluding the day before the vote (Jennings & Wlezien, 2018). In addition, the fact that some countries ban the publication of poll results for periods in the run-up to elections may condition the analysis. Representative polls – with randomly sampled individual – can be expensive and time consuming, even if they historically proved to be quite effective (Jennings & Wlezien, 2018; Wang, Rothschild, Goel, & Gelman, 2015). For this reason, scholars started to look for alternative forecasting methods – such as non-representative polls (Wang et al., 2015) or vote expectation surveys, which Graefe (2014) claimed to be a better and more stable solution. Combined methods also emerged as a possible alternative. For example, Lewis-Beck and Dassonneville (2015) proposed a synthetic approach which integrated predictions based on structural economic variables (static estimation) with aggregated poll results, producing dynamic forecasts. Rothschild (2015) obtained efficient forecasts by aggregating polling, fundamental and prediction market data in a new way .

The era of big data, added complexity to this scenario. On the one hand, scholars and practitioners can now access to multiple data sources which can be used to improve predictions; on the other hand, these data often require costly and difficult operations of crawling, structuring and cleaning, before offering clear signals (Rahm & Do, 2000). However, these new information sources cannot be neglected. The Internet plays a very important role in influencing the behavior of the electorate (Tolbert & McNeal, 2003), with online news being a major information source, followed by social media platforms (Newman et al., 2017). People use the Internet to gather information about political candidates, parties and elections; the larger availability of easily accessible information also stimulates increased electoral participation (Tolbert & McNeal, 2003). The effects produced by online news and social media are still worth of investigation, also considering the more critical views of scholars like Prior (2005, 2013), which lead to the idea that online media, because of selective exposure, mostly reinforce existing attitudes instead of changing opinions. Given this interest for the topic, a more recent trend entails analyzing online big data to see whether they are suitable to make electoral predictions (Linzer & Lewis-Beck, 2015; MacDonald & Mao, 2016). Among different online data sources, Twitter raised the interest of many scholars, also because most tweets are public and relatively easy to collect. Burnap and colleagues (2016) used Twitter to predict the outcome of the UK General Election of 2015, with a model that correctly ranked the first three parties, but had some inaccuracies in determining the number of parliament seats. The methodologies implemented while working with Twitter may vary significantly. The mere counting of the mentions of the names of political parties or candidates does not always produce good results: predictions can be improved with sentiment analysis, by choosing the right search queries and by filtering out those tweets produced by bots or paid users (Ibrahim et al., 2016; Sang & Bos, 2012). Other scholars focused their attention on Google Trends, to predict electoral results by search-query volumes. Mavragani and Tsagarakis (2016) used this kind of data to forecast the Greek Referendum. Similar results could have been obtained by looking at Twitter (Gloor, Fronzetti Colladon, Miller, & Pellegrini, 2016), combining methods and tools of social network and semantic analysis. The analysis of social networks is indeed another promising, but less explored, approach for making political forecasts (Leiter, Murr, Rascón Ramírez, & Stegmaier, 2018). However, even if the big data of social media platforms represent a new source of information, many of the forecasting methods based on these data showed important limitations (Gayo-Avello, Metaxas, & Mustafaraj, 2011). As Huberty showed (2015), many of them fail to produce true forward-looking electoral forecasting and are not able to replace polling in the assessment of voters' intentions.

### 3.1. Brand Importance in Online News and Election Outcomes

Since media reporting deeply influences public opinion, an alternative trend has been to study news to improve political forecasting. News favorability and media coverage of campaign events can indeed favor some candidates over the others, play an important role in shaping political reality (McCombs & Shaw, 1972) and change voters' preferences (Shaw, 1999). Lerman, Gilder, Dredze and Pereira (2008) used computational linguistic to predict the influence of newspaper articles on the public perception of political candidates, achieving improved predictions with respect to political markets. Since the mid-1990s, the Internet has progressively increased its importance as a political campaigning means – extensively used in the United States, for example by Obama in 2008 (Vergeer, 2013). The Web 2.0 can increase both visibility and interactivity of candidates, who can also deliver personalized campaigns. Online news could potentially mitigate the rapid decay of some effects of mass media political communication (Hill, Lo, Vavreck, & Zaller, 2013), as they are indexed by search engines and readable at any moment, also days after their publication. Moreover, online news allow a number of post-read engagement actions, such as social media sharing, commenting, printing and e-mailing article links (Agarwal, Chen, & Wang, 2012). Even if social media play an important role in this scenario, they still have a secondary role when people look for information online (Newman et al., 2017). As a consequence, some scholars attempted to combine the informative power of news and online big data, to make political predictions (Garcia et al., 2018). However, more research is needed in this field. Accordingly, the author is presenting a novel analytical approach to examine the brand importance of political candidates in online news and to make electoral forecasts. Indeed, online news were much less explored than other online data sources, such as Twitter. Another advantage of online news is that they can be timely collected by means of appropriate web crawlers, or by accessing online databases (Leban et al., 2014). Newspaper news, TV and radio broadcasts, on the other hand, are more difficult to collect and examine, also because their content is not always available in a textual form (which is necessary for the calculation of the SBS). Moreover, digital news consumption and impact are increasingly growing, especially among young citizens (Moeller, de Vreese, Esser, & Kunz, 2014), and one could expect that a good part of offline news content is reflected online. Attention should be paid to the phenomenon of online fake news which could sometimes bias the analysis (Newman et al., 2017).

Past research suggested the existence of a relevant link between components of brand equity, such as brand image and awareness, and the success of political candidates and parties (Cwalina & Falkowski, 2015; Guzmán & Sierra, 2009; S. W. Nielsen & Larsen, 2014; Smith, 2001). Indeed, human branding is a relevant research topic in politics (Speed, Butler, & Collins, 2015) and voters' choice has been compared to consumers' choice about commercial brands (Ahmed, Lodhi, & Ahmad, 2017). However, the measurement of the positioning and brand image of political actors can be costly and time consuming, when reliant on the use of questionnaires. The measure of brand importance that is adopted in this study (the SBS, see Section 2) is recent and innovative, and has been specifically developed for the analysis of textual data (Fronzetti Colladon, 2018). It partially reduces the complexity of online big data, enabling a rapid calculation of brand importance in news articles, and it does not require any survey or population sampling. While the construct represented by the SBS is partially new, it is also connected to well-known dimensions of brand equity – such as brand awareness and the study of brand associations (Aaker, 1996; Keller, 1993).

Building on past research which related political candidates and parties to brands (Ahmed et al., 2017; Cwalina & Falkowski, 2015; Guzmán & Sierra, 2009; Scammell, 2007; Smith & French, 2009), the author chose to focus on the construct of brand importance in this work, with the expectation that higher importance can be conductive to greater electoral success. Smith and French (2009) maintained that seeing a party leader on television or reading his/her name, acts as a stimulus, producing activation in citizens' memory. On the one side, visibility and a rich embedding of politicians' names in online news can contribute to their awareness and recall (Hopmann, Vliegenthart, De Vreese, & Albæk, 2010); therefore, citizens that read online news can be primed to think more about those political forces that have a higher SBS. On the other side, it could also be that parties that have a higher standing in the polls receive more attention and media coverage (Rhee, 1996; Rosenstiel, 2005), thus leading to a higher SBS. Both these possibilities – which could be better examined in future research – support the idea of testing the SBS as a promising tool to make political forecasts. The use of the SBS can additionally serve to extend the research on brand importance and political forecasting through a quantitative analysis. To the extent of the author's knowledge, this specific construct, measured through the SBS, has never been used in election forecasting. Moreover, applied in the context of online news, the SBS can offer new insights on the magnitude of the link between online media coverage and political success.

Lastly, since the SBS represents a new construct, it would be difficult to formulate separate hypotheses for the effects produced by each of its dimensions. Consequently, this research is partially exploratory, and the author presents a single general hypothesis that brand importance of political candidates (or party leaders) can be predictive of election outcomes.

## 4. A New Indicator and Methodology

In order to assess the role of brand importance of political parties or candidates in forecasting election results, the author used an adjusted version of the Semantic Brand Score (Fronzetti Colladon, 2018). This novel measure of brand importance, which can be used to analyze big textual data, has been calculated on online news articles.

## 4.1. Data Sample

The online news articles were collected by means of a Python script, interacting with the API system of the Event Registry database. This database contains both mainstream and blog news collected from about 75,000 RSS feeds around the world, producing more than 100,000 articles per day (out of which about 2.74% are written in the Italian language)[1]. This system allows personalized search queries with the possibility to extract news articles that match specific keywords (Leban et al., 2014). Search queries were restricted to online news, excluding blog articles and other sources, such as company press releases.

For this study, the author could collect and analyze the news written in Italian with regard to four voting events in Italy, in the two months preceding them (see Table 1): the Italian Constitutional Referendum in 2016, the Rome Municipal Election in 2016 (first and second round) and the Italian General Election in 2018.

Specific search queries were used to extract all the articles related to each voting event. The author did not only search for the names of political parties and candidates, but also used more general terms related to the voting event (such as "municipal election" + "Rome"). This choice is relevant for the calculation of the SBS, as the author's objective was to capture all the media coverage of each event and of the political forces mentioned before the elections. Therefore, brand importance was assessed considering the overall discourse produced by these large sets of online articles. Consistently, search queries were made of keywords related to each event, together with the names of political forces involved. For example, for the Italian Constitutional Referendum, the author selected the words "referendum",

---

[1] http://eventregistry.org/about, http://newsfeed.ijs.si/, accessed June 5, 2018

"constitutional" and "constitution" (separated by an 'OR' operator) – as these were representative of the main discussion topics, and at least one of them was used in practically all the articles about the Referendum. For this reason, there was no need to add the names of political forces supporting the "Yes" or "No" vote, such as "Renzi". For the Italian General Election 2018, the author used the names of parties and candidates (e.g., "renzi", "salvini", "partito democratico"), together with terms referred to the event, such as "election" or "elections". In addition to party names, the author included their acronyms in the queries, and the nicknames of political candidates when available. However, words related to the ideological leanings of the different parties (e.g. 'socialists') were not used because these terms – which had a significant lower frequency – were sometimes leading to out-of-topic articles and in most cases were redundant, appearing together with at least one of the other keywords. Examples of keywords in this section are in English, but in the search queries the author used the Italian language.

For each event, the articles analyzed dated from about two months up to one day before the vote. Voting days were excluded from the analysis, as many articles contained post-vote information, partial vote counts, or exit-poll results.

| Event | Analysis Period | Voting Day | Analyzed News |
|---|---|---|---|
| Italian Constitutional Referendum 2016 | October 3 - December 3, 2016 | December 4, 2016 | 17,678 |
| Rome Municipal Election 2016 – First Round | April 4 - June 4, 2016 | June 5, 2016 | 4,370 |
| Rome Municipal Election 2016 – Second Round | June 6-18, 2016 | June 19, 2016 | 1,422 |
| Italian General Election 2018 | January 3 - March 3, 2018 | March 4, 2018 | 11,214 |

**Table 1.** Datasets

The data about the voting results were collected from the websites of the Italian government (http://elezionistorico.interno.gov.it/index.php) and of the newspaper 'Il Sole 24 Ore' respectively (http://www.ilsole24ore.com/speciali/2018/elezioni/risultati/politiche/static/italia.shtml).

Lastly, vote intention polls were used as a benchmark to evaluate SBS forecasts (when more poll results were published in a specific week, the author calculated their average). Poll data were collected from the dedicated website of the Italian government (http://www.sondaggipoliticoelettorali.it), where it is mandatory to register political and electoral polls. In general, poll results were not available in the fifteen days before each voting event, as their diffusion is prohibited by the Italian law.

The author chose four events which took place in Italy to preserve the consistency of the analysis with respect to language and local culture. At the same time, he intentionally chose to analyze voting events which were different in terms of voting rules (i.e. a municipal election, a referendum, and a general election). This choice aimed at identifying common patterns in the forecasting methodology.

## 4.2. Text Pre-processing

Following the procedure presented by Fronzetti Colladon (2018), a preliminary step before the calculation of the SBS was the pre-processing of the content of the articles, in order to remove punctuation, special characters and stop-words. The text was subsequently tokenized and each word was converted to lowercase. Lastly, word affixes were removed by using the snowball stemming algorithm of the NLTK Python package (Perkins, 2014). In general, all these operations, as well as the calculation of the SBS, were carried out using the Python programming language and with the help of the functions contained in the packages NLTK and Networkx (Hagberg, Schult, & Swart, 2008; Perkins, 2014).

The analysis has been carried out on the initial 30% text of each article (including the title). This approach reflected the idea that the majority of web users only read a reduced portion of online articles and often stop at the title or at the initial part. How much of an article is read may vary and it depends on many factors, such as the availability of an abstract or the structure and design of webpages (J. Nielsen, 1997, 2008; J. Nielsen & Loranger, 2006). The parts of the text constituting the body of the news collected from the Event Registry were not tagged with the idea of helping the identification of their importance – for example, there was no specific information to distinguish an abstract or lead from the rest of the text (the only distinction was between body and title). Therefore, retaining the initial 30% seemed to be a reasonable approximation – also considering the findings of Nielsen (2008) who maintains that "on the average Web page, users have time to read at most 28% of the words during an average visit". A reduction of the proportion of text to be analyzed has the advantage of significantly decreasing the computation time of the algorithm discussed in the next section.

### 4.3. The Semantic Brand Score

The Semantic Brand Score is a novel indicator, applicable to big text data, to measure the importance of a brand name. This metric and its three components – prevalence, diversity and

connectivity – are well described in the recent work of Fronzetti Colladon (2018) and briefly recalled in the following. According to this indicator, the importance of a brand is high when its name is frequently mentioned, surrounded by heterogeneous textual associations and deeply embedded in a discourse.

The first dimension, prevalence, measures the frequency of use of a specific word (brand) in a set of text documents.

Diversity and connectivity derive from centrality metrics of Social Network Analysis (Wasserman & Faust, 1994) and require the preliminary transformation of the text corpora into a network of word co-occurrences. Accordingly, a collection of text documents can be represented by a network graph, where each node is a word. An arc between two nodes indicates a co-occurrence between two words in the text, within a specific range of words. Arcs can be weighted according to the frequency of co-occurrence of the terms they connect. For example, if the phrase "Happy Holidays!" is repeated ten times, there will be an arc connecting the (stemmed) words 'happi' and 'holiday', whose weight will be 10. Considering the characteristics of the analyzed datasets and the stability analysis presented in the work of Fronzetti Colladon (2018), the co-occurrence threshold for this study was set to 7 words. Moreover, the final graphs were pruned removing those arcs with a weight lower than 2, as probably representing random or unimportant co-occurrences.

 In these networks, diversity is measured by the degree centrality metric (Freeman, 1979) and represents the number of different words co-occurring with a brand name. The higher this number, the higher the number of different textual brand associations.

Connectivity has been originally measured by means of betweenness centrality (Freeman, 1979), which counts how many times a brand name lies in-between the shortest network paths that interconnect all the possible node pairs. This measure has been intended as a proxy of the 'brokerage power' of a brand name, i.e. its ability to connect different parts of the overall discourse (Fronzetti Colladon, 2018). Connectivity, in its first formulation, does not consider arc weights, which however represent an important piece of information to be included in the analysis. The idea is that shortest network paths should take into account the frequency of co-occurrence between pairs of words. Accordingly, links with higher weights should be preferred. When dealing with algorithms that consider arc weight as the distance between two nodes, one can substitute the number of co-occurrences with its reciprocal – so that an arc of original weight 20 will count 1/20 in the computation of network distances. Consequently, the length of a network path is not calculated as the number of arcs it comprises, but as the sum of the weights of those arcs. Following this idea, the author used the algorithm proposed in the work of Brandes (2001) for the calculation of connectivity, thus corresponding to the weighted betweenness centrality of the brand node.

The SBS is not applicable to brands only, but also to any word or set of words, such as the name of political candidates or parties. In this research, brand importance has been measured for candidates' names, party leaders' names, and Yes/No expressions with regard to the Italian constitutional referendum of 2016. When the above-mentioned terms were represented in multiple ways – such as full names of political candidates or just surnames – the analyzed texts were preprocessed, in order to substitute the different expressions of the same name with a single term.

## 5. Results

This section presents the forecasts of electoral results obtained by using the SBS.

### 5.1. Italian Constitutional Referendum 2016

The constitutional referendum was held on December 4, 2016 and strongly promoted by Matteo Renzi, at the time Italy's Prime Minister. Voters were asked to approve a law that amended the Italian Constitution, reforming the composition and powers of the Italian parliament, as well as the division of powers between the State, the regions, and other administrative entities. More than 33,200,000 voters participated in the referendum and only 40.9% of them voted 'Yes'. Consequently, the law was rejected and Renzi handed in his resignation as Prime Minister.

Figure 1 shows the time trends of the SBS for the 'Yes' and 'No' terms, in the 9 weeks preceding the referendum day. As illustrated in Section 4, SBS of each term was calculated as the sum of its standardized prevalence, diversity and connectivity scores. Standardization was carried out by subtracting the mean and then dividing by the standard deviation. Mean and standard deviation were calculated considering all the scores of relevant words in the text. Therefore the measure has not a predefined range, as one would have if normalizing scores between zero and one. Accordingly, one should compare the SBS of a brand with those of competitors or with its historical values. This choice was made by the author, consistently with past research (Fronzetti Colladon, 2018), as other approaches for the calculation of the composite indicator (such as taking the geometric mean) led to a suboptimal forecasting performance. In general, higher scores indicate more important brands, as in the case of the term 'No' which had a higher SBS than the term 'Yes' throughout the entire analysis period. It is also interesting to notice that the name of Matteo Renzi was very strong in the online articles, having a SBS which was superior to the other terms for most of the time and which almost matched the importance of 'Yes' in the last week of analysis. This is also attributable to the fact that Renzi linked the referendum result to his possible resignation, in the case of a 'No' vote. Eventually, this strategy resulted to be counterproductive, since the content of the reform could have found consensus among a majority of voters (Colombo, De Angelis, & Morisi, 2016).

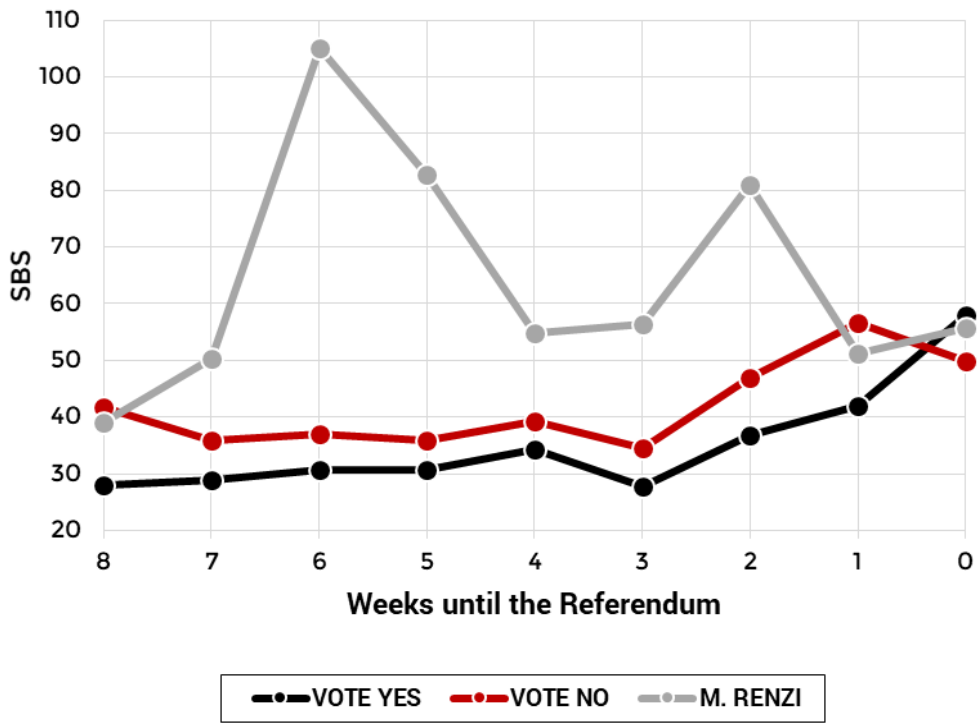

**Figure 1.** Italian Constitutional Referendum 2016 – Semantic Brand Score

Considering the SBS scores reported in Figure 1, the author used the following formula to calculate the relative strength of one term over the other, in order to forecast the proportion of 'Yes' votes.

$$SBSForecast(Yes) = \frac{SBS_{Yes}}{SBS_{Yes} + SBS_{No}}$$

The forecasts are presented in Figure 2, also showing the values that would have been obtained if the SBS dimensions – prevalence, diversity and connectivity – had been considered separately, for example taking the diversity score of "Yes" and dividing it by the sum of the diversity scores of "Yes" and "No". The green line represents the average of vote intention polls; the black line is the final percentage of 'Yes' votes (40.9%).

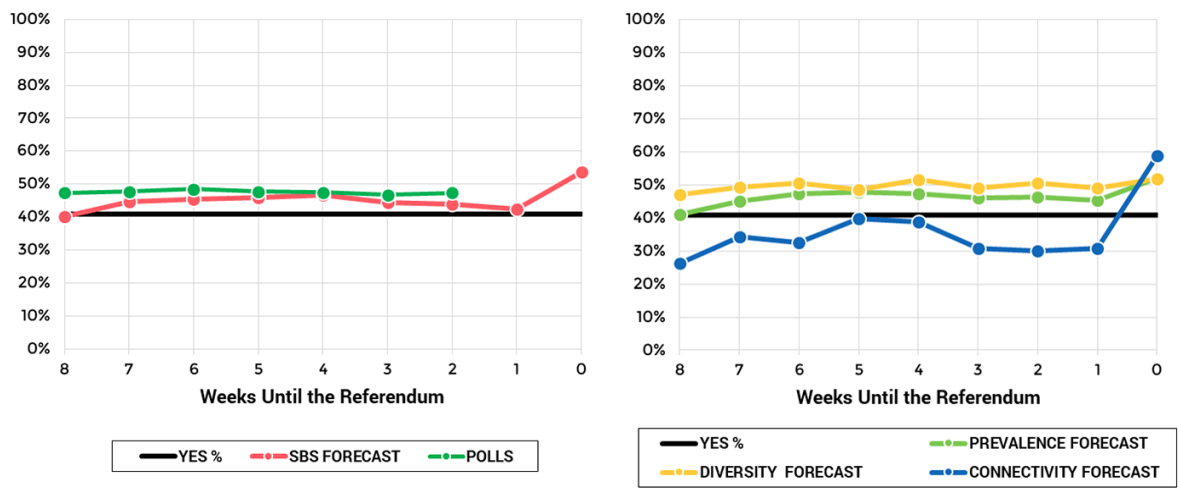

**Figure 2.** Italian Constitutional Referendum 2016 – Forecasts

As the figure shows, the predictions made using the SBS are fairly accurate, especially one week before the referendum. Two weeks before the vote the 'Yes' percentage, forecasted using the SBS, was 43.9% (with 3 percentage points error) and one week before 42.5% (1.6 percentage points error). In both cases the prediction was correctly indicating a victory of the 'No' vote. On the other hand, predictions are not so accurate if considering the week of vote (up to one day before). This result is common to all the case studies presented in this paper, suggesting that the very last week before an election is less useful for this analysis than the one immediately preceding. Moreover, analyzing news in the last week would mean reducing the lead time to one day.

Overall, the results produced by polls tended to overestimate the final percentage of 'Yes' votes and resulted less accurate than the SBS forecasts (with an error of last available polls of 6.45 percentage points). In addition, Figure 2 shows that using the SBS, instead of its separate components, leads to a more stable and accurate forecasting. In this case study, we see a tendency of prevalence and diversity to overestimate the 'Yes' vote and an opposite tendency of connectivity.

## 5.2. Rome Municipal Election 2016

The elections of the Mayor of Rome, the capital city of Italy, were held on June 5, 2016 for the first round and on June 19 for the second round. More than 1,300,000 Roman citizens voted for the first round and 1,185, 280 voted for the second round. The winner turned out to be Virginia Raggi, representative of the political party Movimento 5 Stelle, founded by Beppe Grillo and Roberto Casaleggio in October 2009. This party had an increasing and rapid success also due to its "anti-system stance; its specific conception of democracy; its post-ideological profile; and the innovative elements in its party organization" (Bordignon & Ceccarini, 2015).

### 5.2.1. First Round

In the first round, Virginia Raggi emerged as the favorite candidate, winning about 35% of the votes, followed by Roberto Giachetti (24.91%, representative of the Partito Democratico), Giorgia Meloni (20.62%, representative of the right-wing) and Alfio Marchini (11.00%). The votes obtained by the remaining candidates were never higher than 4.5% and were excluded from the analysis (also because several vote intention polls grouped them under the label 'others', without providing specific forecasts).

The final ranking is replicated exactly by the results shown in Figure 3, if looking at the SBS of each candidate name one week before the election. Examining online news, each candidate's name presented an alternating importance in the two months before the election. However, a general positive trend can be observed for the two candidates who passed to the second round, Virginia Raggi and Roberto Giachetti. Raggi was always more important than Giachetti, which is aligned with the outcomes of the first round.

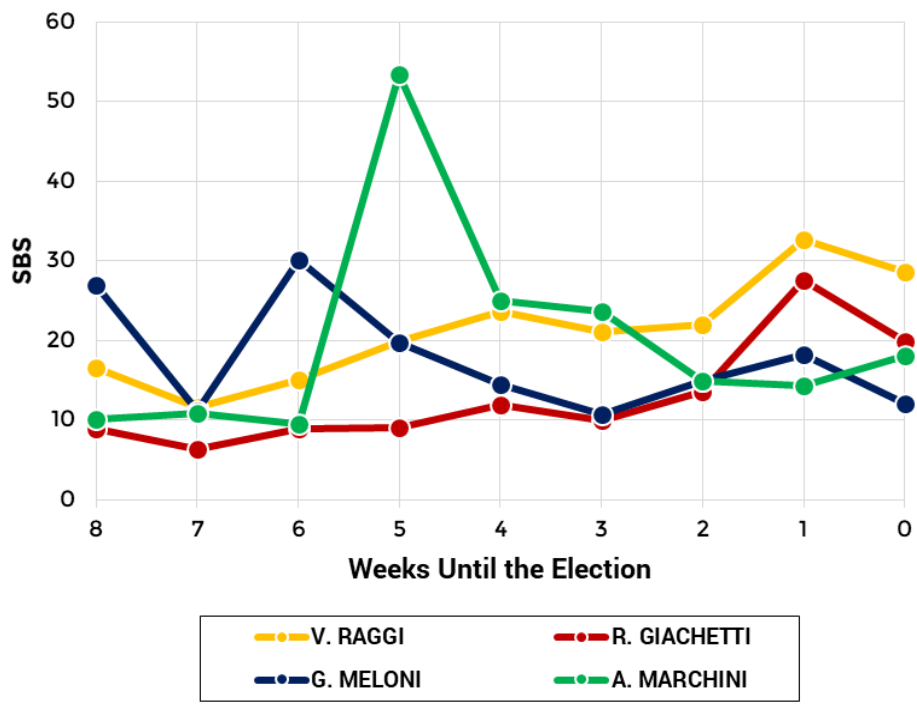

**Figure 3.** First Round Rome Municipal Election 2016 – Semantic Brand Score

With the same procedure of Section 5.1, the SBS scores were used to forecast the percentage of votes obtained by each candidate ($n = 4$):

$$SBSForecast(Candidate_i) = \frac{SBSCandidate_i}{\sum_{i=1}^{n} SBSCandidate_i}$$

The relative strength scores of the four major candidates were adjusted, redistributing the votes obtained by the minor candidates. Table 2 shows the Mean Absolute Percentage Error (MAPE) obtained by the different forecasting approaches. MAPE was calculated using the following formula:

$$MAPE = \frac{\sum_{i=1}^{n} \left| \frac{y_i - \hat{y}_i}{y_i} \right|}{n}$$

where $y_i$ is the actual outcome for $Candidate_i$ and $\hat{y}_i$ is the forecast. Accordingly, the Absolute Percentage Error $\left| \frac{y_i - \hat{y}_i}{y_i} \right|$ is calculated for every candidate and values are subsequently averaged to obtain the MAPE. MAPE is a widely-used and scale-independent metric. It has the potential drawback of putting a heavier penalty on positive errors than on negative errors, as an under-forecast will never contribute more than 100%, but the contribution of an over-forecast is unbounded (Davydenko & Fildes, 2016). However, absolute percentage errors of interest for political forecasting are usually much lower than 100%. MAPE is undefined if an actual value is zero. Fortunately this seems not to be a problem here, as it is highly unlikely that a voting option will take zero percent of votes.

Additionally, the Mean Absolute Error (MAE) was calculated and shown in the tables to provide more complete information:

$$MAE = \frac{\sum_{i=1}^{n} |y_i - \hat{y}_i|}{n}$$

MAE is attractive as it is simple to understand; however, it has several potential limitations, such as being sensitive to outliers and not taking into account relative prediction accuracy. This is an important limitation that suggests that the MAPE is more appropriate to evaluate the forecasts presented in this research. Consider for example two parties which obtain respectively 1 and 40 percent of votes and two corresponding forecasts of 6 and 45 percent. The Absolute Error would be of 5 percentage points in both cases, but the Average Percentage Error would be much higher for the first case than for the second one (500% vs 12.5%), which is more appropriate.

| Date | Weeks Until Election | Polls | | SBS | | Prevalence | | Diversity | | Connectivity | |
|---|---|---|---|---|---|---|---|---|---|---|---|
| | | MAPE | MAE (pp) | MAPE | MAE (pp) | MAPE | MAE (pp) | MAPE | MAE (pp) | MAPE | MAE (pp) |
| W14 2016 | 8 | 8.78% | 2.10 | 51.14% | 12.36 | 47.63% | 10.41 | 52.91% | 11.37 | 89.84% | 20.48 |
| W15 2016 | 7 | | | 54.07% | 10.29 | 48.54% | 9.41 | 45.63% | 9.20 | 92.96% | 15.90 |
| W16 2016 | 6 | 10.48% | 2.90 | 55.53% | 13.90 | 52.77% | 13.33 | 56.02% | 13.21 | 74.26% | 18.74 |
| W17 2016 | 5 | 13.91% | 2.82 | 116.66% | 20.14 | 100.14% | 17.52 | 87.00% | 15.83 | 185.40% | 31.26 |
| W18 2016 | 4 | 26.34% | 4.62 | 63.16% | 10.70 | 51.77% | 8.93 | 51.90% | 8.78 | 121.10% | 20.08 |
| W19 2016 | 3 | 23.92% | 4.30 | 72.38% | 12.12 | 70.77% | 11.90 | 70.26% | 11.88 | 87.92% | 15.12 |
| W20 2016 | 2 | 21.95% | 4.05 | 32.15% | 5.62 | 33.95% | 6.41 | 33.64% | 6.03 | 68.32% | 14.59 |
| W21 2016 | 1 | | | 14.75% | 3.00 | 19.03% | 4.29 | 16.56% | 3.03 | 24.82% | 6.24 |
| W22 2016 | 0 | | | 33.81% | 5.49 | 34.88% | 5.68 | 31.30% | 5.34 | 51.77% | 10.56 |

Note. Pp = percentage points.

**Table 2.** First Round Rome Municipal Election 2016 – SBS Forecast Accuracy

SBS forecasts are more variable and less accurate than polls in the 3 to 9 weeks before the event, with an improving performance as the voting event comes closer. These forecasts seem to be particularly useful at W 21 2016 (once again one week before the election), when the SBS is more informative than the last available polls (MAPE is 14.75%, instead of 21.95%), at a time when further polls are prohibited. At this optimal lead time, SBS also gives better results than its components taken separately. Considering separate dimensions, which at some timeframes had a slightly better performance than the SBS, would have the drawback of only offering a partial view on the brand importance of political candidates (Fronzetti Colladon, 2018). Moreover, which dimension to consider would remain an open question, because the most accurate varies depending on the dataset analyzed and on the time period. In this section case study, prevalence and diversity offered better forecasts than connectivity.

### 5.2.2.  Second Round

In the second round, Virginia Raggi significantly surpassed Roberto Giachetti, winning 67.15% of the total votes and being elected Mayor of Rome. Table 3 shows the forecasts obtained analyzing the two weeks in-between the first and the second round. It is worth noting that the forecast produced through the SBS is very accurate (only 0.58 percentage points error) – much more than those obtained considering prevalence, diversity or connectivity alone. In all forecasts, however, Virginia Raggi is correctly indicated as the winner. In this scenario with only two candidates, the SBS forecast has a significantly higher performance than voting intention polls. This result could also be influenced by the fact that last available polls were administered about two weeks before the first round. Questions made to the interviewees, before the first round, were specifically referred to a hypothetical second round where Raggi would be running against Giachetti.

| | V. Raggi Measure | R. Giachetti Measure | V. Raggi Election Result | Forecast | Absolute Error (pp) | Absolute Percentage Error |
|---|---|---|---|---|---|---|
| Prevalence | 28.74 | 15.90 | 67.15% | 64.38% | 2.77 | 4.12% |
| Diversity | 14.87 | 10.30 | 67.15% | 59.09% | 8.06 | 12.00% |
| Connectivity | 16.03 | 3.76 | 67.15% | 80.99% | 13.84 | 20.62% |
| SBS | 59.64 | 29.95 | 67.15% | 66.57% | 0.58 | 0.87% |
| Average of polls in the two months before the election | | | 67.15% | 57.88% | 9.27 | 13.80% |
| Last available polls | | | 67.15% | 56.89% | 10.26 | 15.27% |

Note. Pp = percentage points.

**Table 3.** Second Round Rome Municipal Election 2016 – Forecast Accuracy

### 5.3. Italian General Election 2018

The 2018 Italian General Election was held on March 4, 2018 and more than 35,000,000 Italians voted. The bicameral Italian Parliament is composed of two houses, the Chamber of Deputies and the Senate of the Republic. The percentage of votes obtained by the major political parties in the two houses were extremely similar[2]. In this experiment, the author used the results of the Chamber of Deputies election, as the number of people who were enabled to vote was higher than that for the Senate (all those aged 18 years or more, instead of 25 years or more): Movimento 5 Stelle, 32.66%; Partito Democratico 18.72%; Lega, 17.37%; Forza Italia, 14.01%; Fratelli D'Italia, 4.35%; Liberi e Uguali, 3.39%. The other political parties which obtained less than 3% of votes were excluded from the experiment, with a consequent adjustment of the above mentioned percentages.

---

[2]http://www.ilsole24ore.com/speciali/2018/elezioni/risultati/politiche/static/italia.shtml?refresh_ce=1, accessed June 5, 2018.

Figure 4 shows the SBS of each party representative's name – Luigi Di Maio (Movimento 5 Stelle), Matteo Renzi (Partito Democratico), Matteo Salvini (Lega), Silvio Berlusconi (Forza Italia), Giorgia Meloni (Fratelli d'Italia), Pietro Grasso (Liberi e Uguali) – in the eight weeks preceding the vote. For the Movimento 5 Stelle, the name of Beppe Grillo was also included in the analysis (and combined with Di Maio), as he is recognized as one of the party leaders and he is often mentioned in the news about the movement (Bordignon & Ceccarini, 2015). Also in this case, results at one week before the vote are the most informative, i.e. the ranking of political parties are closer to the election outcomes (with the exception of Silvio Berlusconi, whose brand importance was higher than the others, even if this translated in a less than proportional share of votes). It is also worth noting that SBS trends of main political forces were less stable than minor parties.

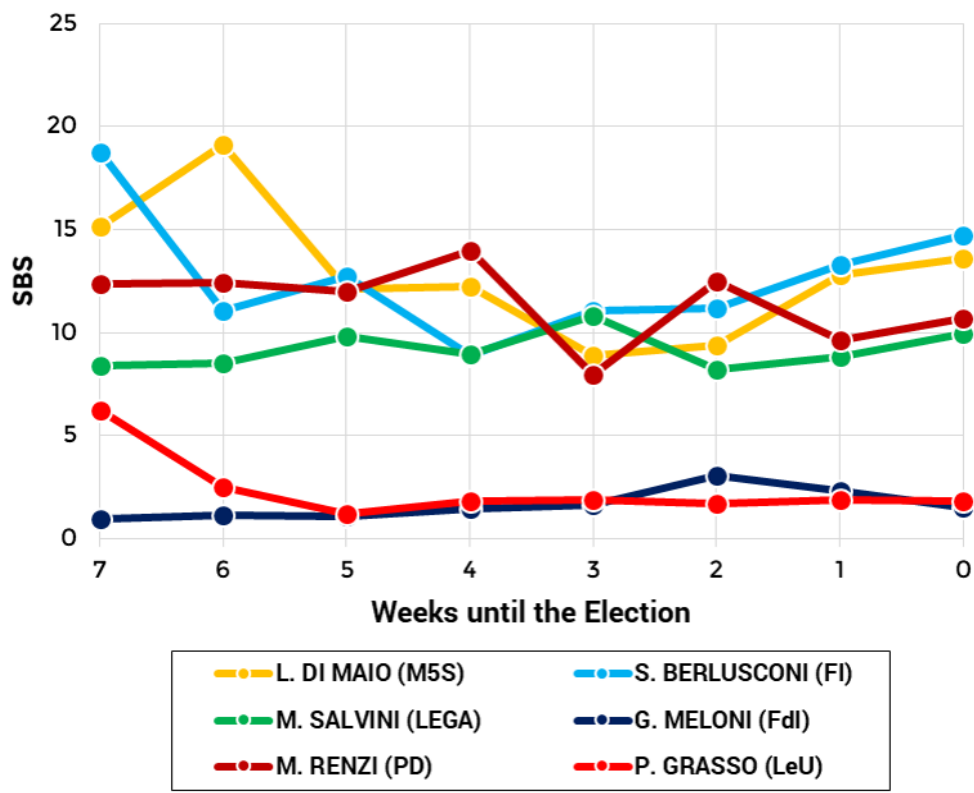

**Figure 4.** Italian General Election 2018 – Semantic Brand Score

The forecasts for this scenario are more complex than the two presented in Sections 5.1 and 5.2, as percentages of votes for six political parties have to be predicted. The formula used is similar to those of the other case studies:

$$SBSForecast(Party_i) = \frac{SBSPartyLeader_i}{\sum_{i=1}^{n} SBSPartyLeader_i}$$

Table 4 shows the MAPE of the different forecasting approaches. Similarly to the case presented in Section 5.2.1, there is more variation in the forecasts produced by the SBS and its components, with respect to vote intention polls. However, the SBS forecast obtained one week before the election is the one with the smallest error (MAPE = 19.76%) and outperforms the MAPE of last available polls (25.63%).

| | | Polls | | SBS | | Prevalence | | Diversity | | Connectivity | |
|---|---|---|---|---|---|---|---|---|---|---|---|
| Date | Weeks Until Election | MAPE | MAE (pp) | MAPE | MAE (pp) | MAPE | MAE (pp) | MAPE | MAE (pp) | MAPE | MAE (pp) |
| W2 2018 | 7 | 29.48% | 3.49 | 66.10% | 7.06 | 69.22% | 7.55 | 66.68% | 6.70 | 59.60% | 7.65 |
| W3 2018 | 6 | 29.97% | 3.74 | 23.46% | 2.50 | 20.26% | 2.31 | 29.41% | 3.63 | 59.91% | 8.97 |
| W4 2018 | 5 | 29.66% | 3.68 | 35.25% | 5.07 | 33.50% | 5.64 | 37.26% | 5.38 | 53.96% | 5.21 |
| W5 2018 | 4 | 26.95% | 3.58 | 22.24% | 4.10 | 21.96% | 4.04 | 25.32% | 4.68 | 63.07% | 6.68 |
| W6 2018 | 3 | 26.10% | 3.38 | 31.95% | 5.95 | 29.36% | 5.38 | 30.79% | 5.39 | 103.10% | 14.52 |
| W7 2018 | 2 | 25.63% | 3.36 | 29.88% | 5.73 | 30.66% | 5.66 | 32.64% | 5.55 | 66.36% | 8.95 |
| W8 2018 | 1 | | | 19.76% | 3.97 | 24.71% | 4.35 | 20.97% | 4.17 | 67.62% | 8.33 |
| W9 2018 | 0 | | | 26.42% | 4.22 | 23.91% | 3.86 | 26.88% | 4.40 | 68.50% | 7.90 |

Note. Pp = percentage points.

**Table 4.** Italian General Election 2018 – SBS Forecast Accuracy

This experiment was also replicated measuring the SBS of each party name, as an alternative to the SBS of party leaders. In addition, party names were combined with those of their representatives. However, none of these variations to the original methodology led to better results.

## 5.4. Summary of Results

Table 5 presents a summary of the results discussed in the previous sections, giving evidence to potential of the SBS to make political forecasts, and more in general to the possible association of brand importance with election outcomes.

**Italian General Election 2018**

| Measure | Time | MAPE | MAE (pp) |
|---|---|---|---|
| SBS | Lag 1 Week | 19.76% | 3.97 |
| Prevalence | Lag 1 Week | 24.71% | 4.35 |
| Diversity | Lag 1 Week | 20.97% | 4.17 |
| Connectivity | Lag 1 Week | 67.62% | 8.33 |
| Polls | Last Available (Lag 2 Weeks) | 25.63% | 3.36 |

| Political Party | Election Result | Adjusted Election Result | SBS Forecast (Lag 1 Week) | Absolute Error (pp) | Absolute Percentage Error | Real Rank | Forecasted Rank |
|---|---|---|---|---|---|---|---|
| Movimento 5 Stelle | 32.66% | 36.09% | 26.26% | 9.83 | 27.23% | 1 | 2 |
| Partito Democratico | 18.72% | 20.69% | 19.74% | 0.94 | 4.55% | 2 | 3 |
| Lega | 17.37% | 19.19% | 18.13% | 1.07 | 5.55% | 3 | 4 |
| Forza Italia | 14.01% | 15.48% | 27.26% | 11.78 | 76.12% | 4 | 1 |
| Fratelli d'Italia | 4.35% | 4.81% | 4.73% | 0.08 | 1.65% | 5 | 5 |
| Liberi e Uguali | 3.39% | 3.75% | 3.87% | 0.13 | 3.43% | 6 | 6 |

**Rome Municipal Election 2016 - First Round**

| Measure | Time | MAPE | MAE (pp) |
|---|---|---|---|
| SBS | Lag 1 Week | 14.75% | 3.00 |
| Prevalence | Lag 1 Week | 19.03% | 4.29 |
| Diversity | Lag 1 Week | 16.56% | 3.03 |
| Connectivity | Lag 1 Week | 24.82% | 6.24 |

| Polls | Last Available (Lag 2 Weeks) | 21.95% | 4.05 | | | | |
|---|---|---|---|---|---|---|---|
| **Political Candidate** | **Election Result** | **Adjusted Election Result** | **SBS Forecast (Lag 1 Week)** | **Absolute Error (pp)** | **Absolute Percentage Error** | **Real Rank** | **Forecasted Rank** |
| V. Raggi | 35.26% | 38.41% | 35.19% | 3.22 | 8.40% | 1 | 1 |
| R. Giachetti | 24.91% | 27.14% | 29.69% | 2.55 | 9.40% | 2 | 2 |
| G. Meloni | 20.62% | 22.46% | 19.69% | 2.77 | 12.36% | 3 | 3 |
| A. Marchini | 11.00% | 11.98% | 15.43% | 3.45 | 28.85% | 4 | 4 |

<div align="center"><strong>Rome Municipal Election 2016 - Second Round</strong></div>

| **Measure** | **V. Raggi Election Result** | **Forecast** | **Absolute Error (pp)** | **Absolute Percentage Error** |
|---|---|---|---|---|
| SBS | 67.15% | 66.57% | 0.58 | 0.87% |
| Prevalence | 67.15% | 64.38% | 2.77 | 4.12% |
| Diversity | 67.15% | 59.09% | 8.06 | 12.00% |
| Connectivity | 67.15% | 80.99% | 13.84 | 20.62% |
| Polls | 67.15% | 56.89% | 10.26 | 15.27% |

<div align="center"><strong>Italian Constitutional Referendum 2016</strong></div>

| **Measure** | **Time** | **Yes vote** | **Forecast** | **Absolute Error (pp)** | **Absolute Percentage Error** |
|---|---|---|---|---|---|
| SBS | Lag 1 Week | 40.90% | 42.50% | 1.60 | 3.91% |
| Prevalence | Lag 1 Week | 40.90% | 45.21% | 4.31 | 10.54% |
| Diversity | Lag 1 Week | 40.90% | 49.13% | 8.23 | 20.12% |
| Connectivity | Lag 1 Week | 40.90% | 30.77% | 10.13 | 24.76% |
| Polls | Last Available (Lag 2 Weeks) | 40.90% | 47.35% | 6.45 | 15.76% |

* p < .05; **p < .01; ***p < .001. Pp = percentage points.

**Table 5.** Summary of Results

The forecasts made by calculating the SBS on online news were fairly accurate in all the three scenarios considered in this paper – a referendum, a municipal election and a general election. If forecasting at longer horizons, predictions made by using the SBS were less accurate and stable – and, in cases with more than two voting options, they were outperformed by vote intention polls. However, accuracy of SBS forecasts significantly increased, in all the case studies, one week before the vote: at this optimal lead time, the MAPE of SBS predictions was always better than the one of last available polls. The highest performance could be achieved in cases where the number of options was reduced (a Yes/No or a runoff voting). In these scenarios, SBS forecasts performed better than polls, even at longer horizons.

In the most complex scenario – where the relative strength of six parties had to be predicted – the maximum absolute error, one week before the vote, was of 11.4 percentage points (APE = 76.12%), overestimating the results of Forza Italia, the party led by Silvio Berlusconi. By contrast, vote shares of Movimento 5 Stelle were underestimated of 9.83 percentage points (APE = 27.23%). Forecasts for the remaining four parties where much more accurate. Larger errors might be attributable to the fact that Berlusconi – whose companies control several TV channels, newspapers and magazines (Mazzoleni, 2016) – had a larger exposure on online news than Di Maio; this produced a higher SBS and a shift in forecasted rankings, but eventually did not translate into more votes.

## 6. Discussion and Conclusions

This work presented a novel analysis of big data extracted from online news, where the main purpose was to calculate brand importance of political forces and use it to make electoral forecasts. To accomplish that, brand importance has been measured through a recently developed indicator, the Semantic Brand Score (Fronzetti Colladon, 2018). The obtained results suggest the existence of a link between brand importance of political candidates and their success.

Forecasts were intentionally tested considering three different voting systems – a municipal election, a referendum and a general election – and produced consistent results. As one could expect, the percentage error was lower in the cases where the number of available voting options was lower. For example, for the Italian constitutional referendum, the SBS could predict the voting results with only an error of only 1.6 percentage points (error of last available polls was 6.45 percentage points). In the case of the Italian General Election of 2018, the MAPE at one-week before the vote was 19.76% (MAPE of last available polls was 25.63%), with the bigger error made for the party of Silvio Berlusconi. This could suggest the necessity of an adjustment of results when a political candidate is also a media tycoon (Mazzoleni, 2016), who has more media exposure, which does not necessarily translates into more votes. In addition, future research could explore the existence of factors which moderate the relationship between brand importance and voting intentions.

SBS forecasts showed to have an optimal lead time of one week. These forecasts were more stable in scenarios with two voting options and less stable in the cases with a higher number of candidates. However, predicted rankings were almost always correct and the SBS could produce better forecasts than its separate components (probably because only combined they can capture the full construct of brand importance). Vote intention polls were offering more stable predictions (in the cases with 4 or 6 voting options), but last available polls were always outperformed by the SBS forecasts at its optimal lead time. It is important to notice that the aim of this research is not to prove that SBS forecasts are better than vote intention polls. Indeed, in the weeks when polls were available, their prediction errors were often lower than those of SBS forecasts, or similar to them. SBS forecasts always outperformed polls only in the case of the Italian Constitutional Referendum, with voting options restricted to two ('Yes' and 'No'). A fairer comparison would be possible if polls were not banned one week before elections, at the time when SBS is most informative. Therefore, information coming from SBS forecasts complements vote intention polls, especially during the banning period.

In general, it is difficult to speculate on the reasons behind the SBS optimal lead time of one week – which does not seem to be a coincidence as, in all case studies, forecasts proved to be worse in the weeks before and in the week of the vote. One reason could be that online articles may require some time, after they are published, to spread and get indexed by search engines, thus reaching more voters. According to this (untested) idea, the articles published in the week before the voting week could be those with a better indexing. In addition, online news could produce an effect which lasts longer than a TV newscast or a tweet (as their content is searchable, easier to access a second time, or share via mobile apps like WhatsApp). Furthermore, the week with the best SBS forecasts coincides with the beginning of polls banning, which in Italy starts 15 days before elections. During that week, the phenomenon of 'horse race' reporting (Rosenstiel, 2005) may be lower and online news may be less influenced by poll results. Moreover, when polls are available, their media coverage could be biased and favor results which are more exciting and newsworthy (Searles, Ginn, & Nickens, 2016), ultimately creating a disturbance in the signal captured by the SBS. In general, the beginning of poll banning seems beneficial to the accuracy of SBS forecasts and the hypothetical reasons here outlined are certainly worth further investigation.

Weekly forecasts were chosen to reduce the variability of SBS scores, which depend on the news published each day. Daily forecasts are possible and the author tested them, but they produced less accurate and stable results. Using a weekly timeframe has other advantages, such as an easier comparison with polls (that are not published every day). Moreover, one could expect that the hypothetical effect of online news on voters' intentions comes from the combined exposure to several days of news. Indeed, online news are probably better indexed by search engines and online aggregators after some time, so their maximum diffusion is difficult to attribute to specific days. An interesting analysis could be carried out, if information about web traffic for each news was available to the analyst. It is worth noting that the SBS calculated over a week-period is different from the one obtained by averaging daily scores.

In the analysis of the Italian General Election 2018, measuring brand importance of party leader names led to much more accurate forecasts than considering brand importance of party names. This is consistent with the study of Speed et al. (2015) – which maintained the importance of human branding in political marketing – and with the idea that reading the name of a party leader (and perhaps seeing his pictures) in online news could stimulate and activate voters' memory (Smith & French, 2009).

The methodology presented in this paper can be intended as a first step in the improvement of existing forecasting models, which could incorporate the information coming from the SBS. This information is not meant to replace polls, but could complement them, especially when the voting day comes closer and in countries where polls are banned in the days/weeks before the election. In addition, this new approach has the advantage of being easily implementable for a 'low-cost' continuous monitoring of the importance of political brands, as it does not require costly and time-consuming surveys. Moreover, a preliminary analysis showed that the assessment of brand importance can be successfully limited to the initial 30% of each article, reducing the computation time of the SBS algorithm. This approach is aligned with the idea that a large part of web users only read the beginning of online articles (J. Nielsen, 2008). One could also compare the forecasts obtained by considering the 30% and the 100% of news texts: this has been done in a preliminary analysis of the presented case studies and led to no better results.

Online news at least partially mirror the discourse conveyed on TV, newspapers and other media. However, future research is advocated for to explore the contribution of the SBS as a predictor of political results, while considering other media sources or social media data, other languages or world regions. The SBS could be combined with sentiment analysis in future studies, which could additionally aim to increase the forecasting lead of this indicator. Even if the SBS showed to be a promising and valid measure across different voting systems, there are past models which could make predictions with longer leads. Therefore, future research could consider an integration of different techniques and parallel tests on the same sets of events. Brand importance is a specific construct, which could be integrated with different measuring systems of brand image and brand equity.

This research has some limitations that partially rely on the use of online news, without considering other media sources – such as newspapers, social media, blogs or TV broadcasts. Future research could compare the author's findings with those obtained from the analysis of other sources. In addition, not every website has the same importance and some webpages have much more resonance than others. Therefore, future analysts could use a weighting system to attribute more importance to some articles – considering, for example, the average number of visitors of each website. Lastly, the analysis could be repeated, classifying the most important parts of each article, without limiting the selection to the first 30% of words; an intelligent crawler could distinguish between title, subtitle, news lead or abstract (if present) and subsequent text parts. Future research could also test different approaches in the selection of the articles for each voting event.

It is important to notice that score variations of the SBS of political candidates' names are per se informative, even before using them for forecasting purposes. However, the author has no ambition to comment on or interpret the reasons behind the presented political outcomes. This is something left to dedicated research with a different scope.

Election forecasting can support both academic and public purposes. Parties and their representatives can adjust media and communication strategies, or change the target of their campaigns. Scholars, on the other hand, can benefit from new forecasting methodologies to extend their research (Lewis-Beck, 2005). This work extends the research about brand importance, the SBS and its applications. The author proposed an adjusted way of calculating connectivity, by using weighted betweenness centrality (Brandes, 2001). Future studies could also test other variations of the original betweenness centrality algorithm (Freeman, 1979), including for example flow betweenness (Freeman, Borgatti, & White, 1991), or other generalizations (Kivimäki, Lebichot, Saramäki, & Saerens, 2016; Opsahl, Agneessens, & Skvoretz, 2010).

The new approach illustrated in this paper extends the research on political forecasting by news analysis (Garcia et al., 2018; Lerman et al., 2008), through methods and tools of social network analysis and text mining. The author's findings are important not only in terms of forecasting, they also give evidence to the influence that online news can have on voting intentions (Shaw, 1999; Vergeer, 2013). Indeed, the question whether the online news are better for revealing or influencing electoral results remains open.